\documentclass[letterpaper,aps,pre,twocolumn,showcaps,superscriptaddress,showpacs,amsmath,amssymb,nobalancelastpage]{revtex4}
\usepackage{graphicx}
\usepackage[usenames]{color}

\newcommand{\dd}{\mathrm{d}}

\newcommand{\mean}[1]{\langle #1 \rangle}
\newcommand{\Int}[1]{\int\dd #1\;}
\newcommand{\IInt}[3]{\int_{#2}^{#3}\dd #1\;}
\newcommand{\Path}[1]{\int[\dd#1]\;}

\newcommand{\gam}{\gamma}

\newcommand{\kap}{\kappa}
\newcommand{\lam}{\lambda}

\newcommand{\ra}{\rightarrow}

\newcommand{\im}{\text{i}}
\newcommand{\kT}{k_\text{B}T}
\newcommand{\x}{\mathbf r}
\newcommand{\q}{\mathbf q}
\newcommand{\Ha}{\mathcal H}
\newcommand{\Hi}{\Ha_\text{i}}
\newcommand{\Hr}{\Ha_\text{r}}
\newcommand{\Z}{\mathcal Z}
\newcommand{\F}{\mathcal F}

\newcommand{\Fex}{F^\text{ex}}
\newcommand{\Q}{\mathcal Q}
\newcommand{\Ns}{N_\text{S}}
\newcommand{\Nl}{N_\text{L}}
\newcommand{\eb}{\epsilon_\text{b}}
\newcommand{\rgh}{\xi_\perp}    
\newcommand{\rghd}{\xi_{\perp,\text{d}}}
\newcommand{\grgh}{\bar\xi_\perp}
\newcommand{\cor}{\xi_\parallel} 

\DeclareMathOperator{\kei}{kei}

\begin{document}

\title{Effective free energy for pinned membranes}
\author{Thomas Speck}
\affiliation{Department of Chemistry, University of California, Berkeley,
  California 94720, USA}
\affiliation{Chemical Sciences Division, Lawrence Berkeley National
  Laboratory, Berkeley, California 94720, USA}

\begin{abstract}
  We consider membranes adhered through specific receptor-ligand
  bonds. Thermal undulations of the membrane induce effective interactions
  between adhesion sites. We derive an upper bound to the free energy that is
  independent of interaction details. To lowest order in a systematic
  expansion we obtain two-body interactions which allow to map the free energy
  onto a lattice gas with constant density. The induced interactions alone are
  not strong enough to lead to a condensation of individual adhesion sites. A
  measure of the thermal roughness is shown to depend on the inverse square
  root of the density of adhesion sites, which is in good agreement with
  previous computer simulations.
\end{abstract}


\maketitle


Living cells interact with their environment through the cell membrane, a
bilayer of phospholipids consisting of hydrophobic tails and hydrophilic
heads. Embedded integral proteins, which make up a large portion of the
membrane, are involved in a range of processes such as ion passage and cell
signaling. Through the formation of receptor-ligand bonds certain proteins
such as cadherins, selectins, and integrins are involved in cell adhesion,
where one cell binds to another cell, to an extracellular matrix, or to a
surface~\cite{beckerle}. Extensive studies have been conducted on cell mimetic
model systems using a variety of
techniques~\cite{albe97,tana05,moss07,smit10a}.

The proteins embedded into the membrane can move rather freely together with
the lipids. The lateral dynamics and diffusion properties of embedded integral
proteins have been studied both theoretically~\cite{reis05,brow08} and
experimentally~\cite{kaiz04}. Protein dynamics is influenced by the
cytoskeleton, geometric constraints, and the thermal undulations of the
membrane itself. Adhesion sites (i.e., single receptor-ligand bonds) in cells
are not randomly distributed but form focal adhesions which are several square
microns in area~\cite{geig01}. Since the bending of the membrane requires
energy one would imagine that such a clustering of adhesion sites is
preferential. However, computer simulations of confined
membranes~\cite{krob07,weik07,fara10,weil10} indicate that fluctuation-induced
interactions between adhesion sites are short-ranged and insufficiently strong
to lead to thermodynamically stable clusters, quite in contrast to earlier
predictions~\cite{brui94,zuck95}. The aim of this Rapid Communication is to
derive an effective and accurate expression for the interaction free energy of
pinned membranes through a variational approach. Membrane-induced interactions
between adhesion sites are indeed found to be rather weak, confirming that
active processes and inter-protein interactions are primarily responsible for
the formation of adhesion domains.


\begin{figure}[b]
  \centering
  \includegraphics{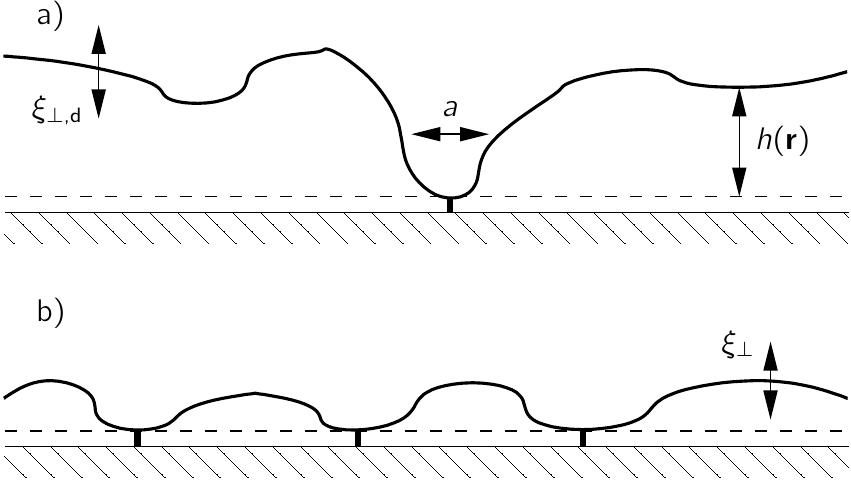}
  \caption{Sketch of a membrane with instantaneous height profile $h(\x)$
    moving above a substrate. The membrane is pinned down through adhesion
    centers with effective size $a$. a) In the presence of only a few centers
    the thermal roughness away from the adhesion centers is $\rghd$. b) For a
    large number of adhesion centers the membrane is pulled closer to the
    substrate. This reduces the thermal roughness $\rgh<\rghd$ due to
    interactions of the membrane with the substrate.}
  \label{fig:mem}
\end{figure}

We consider a tension-less, almost flat membrane moving close to a substrate
in a fluid at temperature $T$. The membrane height profile $h(\x)$ is given in
the Monge representation with $\x=(x,y)$. The projected area is $L^2$ and for
simplicity we employ periodic boundary conditions. In this geometry the
membrane can be thought of as a two dimensional sheet, where the thermal
fluctuations are governed by the Helfrich energy~\cite{helf78}. We focus on
the bending energy
\begin{equation}
  \label{eq:H0}
  \Ha_0[h(\x)] = \Int{^2\x} \frac{\kap}{2}[\nabla^2h(\x)]^2
\end{equation}
as the dominant contribution with bending rigidity $\kap$. In the presence of
$N$ adhesion centers at locations $\{\x_i\}$ pinning the membrane to the
substrate the partition sum reads
\begin{equation}
  \label{eq:ZN}
  \Z_N(\{\x_i\}) \equiv \Path{h(\x)} e^{-\beta(\Ha_0+\Hi)} 
  \prod_{i=1}^N \ell\delta(h(\x_i))
\end{equation}
with $\beta\equiv(\kT)^{-1}$. The path integral sums over all accessible
membrane configurations $h(\x)\geqslant0$. The (small) length $\ell$ is
related to the flexibility of bonds and quantifies the fluctuations of the
membrane at the adhesion sites (one should think of the $\delta$-functions as
limits of Gaussians). Van der Waals forces, the repulsion due to the formation
of an electrostatic double-layer, and steric forces give rise to an
interaction energy $\Hi[h(\x)]$ between substrate and membrane, see
Fig.~\ref{fig:mem}. While explicit expressions can be found for these
forces~\cite{israel}, here we pursue a more general strategy that does not
depend on specific details.


We expand the height profile $h(\x)$ into Fourier modes,
\begin{equation*}
  h(\x) = \sum_\q h_\q e^{-\im\q\cdot\x}, \quad
  h_\q = \frac{1}{L^2} \Int{^2\x} h(\x) e^{\im\q\cdot\x},
\end{equation*}
where $h_\q^\ast=h_{-\q}$. In particular, $h_0$ is the height averaged over
the area of the membrane. The Helfrich Hamiltonian~\eqref{eq:H0} becomes
\begin{equation*}
  \Ha_0 = \frac{\kap L^2}{2}\sum_\q q^4|h_\q|^2 
  = \frac{1}{2}\sum_{\q\in\Q}2\kap L^2q^4[h_\q'^2+h_\q''^2],
\end{equation*}
where we split $h_\q=h_\q'+\im h_\q''$ into real and imaginary part and
$q\equiv|\q|$. We sum only over independent wave vectors, i.e., for a given
$\q$ the set $\Q$ does not contain $-\q$. Also, $q=0$ is excluded from the
sum.

Due to the pinning the membrane is pulled towards the substrate. The
interactions of the membrane with the substrate lead to a 'confinement' of
fluctuations. For membranes placed between walls the interaction energy is
often expanded into a quadratic function of the height. However, in principle
we can do better through introducing a quadratic reference energy $\Hr$ that
acts on every height mode independently. Employing the Jensen inequality we
know that the free energy is bounded from above as~\cite{chandler}
\begin{equation}
  \label{eq:F:var}
  \F_N \leqslant \tilde F_N + \mean{\Delta\Ha} \equiv F_N
\end{equation}
with $\F_N\equiv-\beta^{-1}\ln\Z_N$. Here, $\Delta\Ha\equiv\Hi-\Hr$ and
$\mean{\cdots}$ denotes the average with respect to the reference system with
partition sum $\tilde Z_N$ and free energy $\tilde
F_N\equiv-\beta^{-1}\ln\tilde Z_N$. By minimizing the function $F_N$ with
respect to a set of free parameters we obtain an approximation of the true
free energy $\F_N(\{\x_i\})$. In this work we choose an intermediate
description
\begin{equation}
  \label{eq:H:ref}
  \Hr \equiv \frac{1}{2}\gam_0(h_0-\bar h)^2 +
  2L^2\gam \sum_{\q\in\Q}[h_\q'^2+h_\q''^2]
\end{equation}
with free parameters $\{\gam_0,\bar h,\gam\}$. The total energy then reads
\begin{equation}
  \label{eq:H:tot}
  \beta(\Ha_0+\Hr) = \frac{1}{2}\hat\gam_0(h_0-\bar h)^2
  + \frac{1}{2}\sum_{\q\in\Q}\hat\gam_q[h_\q'^2+h_\q''^2]
\end{equation}
with $\hat\gam_0\equiv\beta\gam_0$ and $\hat\gam_q\equiv 2L^2\beta(\gam+\kap
q^4)$.


Before we calculate the reference partition sum $\tilde Z_N$ let us assume we
had performed the minimization of $F_N$ in the presence of the adhesion
centers with constraints $h(\x_i)=0$. We hypothetically remove the constraints
but keep the values for $\gam_0$, $\bar h$, and $\gam$ we had obtained as a
result of the minimization. We define the thermal roughness $\rgh$ through
\begin{equation}
  \label{eq:rgh:def}
  \rgh^2 \equiv \mean{[h(\x)-h_0]^2}_0 = \sum_{\q\neq0}\mean{|h_\q|^2}_0 =
  4\sum_{\q\in\Q} \frac{1}{\hat\gam_q},
\end{equation}
which is independent of $\gam_0$ and $\bar h$. The brackets $\mean{\cdots}_0$
denote the average involving the energy Eq.~\eqref{eq:H:tot} in the absence of
constraints. For the sake of simplicity and for sufficiently large $L$ we
evaluate the sum over wave vectors $\q$ through a two-dimensional integral
over the half-plane with the well-known result
\begin{equation}
  \label{eq:rgh}
  \rgh \approx 
  \frac{\pi L^2}{(2\pi)^2} \IInt{q}{2\pi/L}{\infty} 
  q\frac{4}{\hat\gam_q}
  \approx \cor/\sqrt{8\beta\kap} \leqslant \rghd
\end{equation}
in the limit $L\gg\cor$. Here, $\cor\equiv(\kap/\gam)^{1/4}$ is the lateral
decay length over which membrane fluctuations are correlated. The thermal
roughness is bounded by its value $\rghd\approx L/\sqrt{16\pi^3\beta\kap}$ for
a free, detached membrane with $\gam=0$. Note that the thermal roughness
$\grgh$ obtained by averaging over the projected area including the adhesion
sites is greater than $\rgh$.


We integrate over the membrane height fluctuations in the presence of $N$
adhesion centers at positions $\{\x_i\}$~\cite{zuck95,spec10b}. The reference
partition sum including the constraints reads
\begin{multline}
  \label{eq:6}
  \tilde Z_N = (\ell/2\pi)^N \Path{h(\x)} e^{-\beta(\Ha_0+\Hr)}
  \Int{\lam_1\cdots\dd\lam_N} \\ \times \exp\left\{
    \im h_0\sum_{i=1}^N \lam_i + \im \sum_{\q\in\Q}[c_\q'h_\q' +
    c_\q''h_\q''] \right\}
\end{multline}
with coefficients
\begin{equation*}
  c_\q' \equiv 2\sum_{i=1}^N \lam_i\cos\q\cdot\x_i, \qquad
  c_\q'' \equiv 2\sum_{i=1}^N \lam_i\sin\q\cdot\x_i.
\end{equation*}
We have replaced the $\delta$-constraints [Eq.~\eqref{eq:ZN}] by integrals
over a set of auxiliary variables $\{\lam_i\}$. We assume that the excluded
volume constraint is represented by the interaction energy $\Hi$, and
therefore by $\Hr$. Performing the integrations we obtain
\begin{equation}
  \label{eq:Z:eff}
  \tilde Z_N \sim \tilde Z_0 [\det m/\ell^{2N}]^{-1/2}, \quad
  \tilde Z_0 = \prod_{\q\in\Q} \frac{1}{\hat\gam_q},
\end{equation}
where we have dropped the sub-extensive contribution due to the integration
over $h_0$. The symmetric $N\times N$ matrix $m$ has components $m_{ij}\equiv
m(|\x_i-\x_j|)$, where
\begin{equation}
  \label{eq:m}
  m(r) \equiv 4\sum_{\q\in\Q} \frac{\cos\q\cdot\x}{\hat\gam_q}
  \approx \rgh^2 u(r/\cor)
\end{equation}
describes the membrane-mediated interactions. The spatial correlations between
height fluctuations and therefore the effective lateral interactions between
adhesion sites is short ranged with potential $u(x)\equiv-(4/\pi)\kei_0(x)$,
where $\kei_0(x)$ is a Kelvin function~\cite{abramowitz}. The effective
interactions decay on the single length scale $\cor$ that also determines the
thermal roughness $\rgh$ [Eq.~\eqref{eq:rgh}].


For the calculation of $F_N$ [Eq.~\eqref{eq:F:var}] we first determine the
contribution
\begin{equation}
  -\ln\tilde Z_0 = \sum_{\q\in\Q} \ln\hat\gam_q = L^2/(8\cor^2),
\end{equation}
which for small $\cor$ captures the loss of entropy due to a large confinement
of fluctuations. To establish $F_N$ as a rigorous upper bound to the free
energy $\F_N$ we impose
\begin{equation}
  \label{eq:H:up}
  \beta\mean{\Delta\Ha} \leqslant -\beta\mean{\Hr}_0 = -L^2/(4\cor)^2.
\end{equation}
This bound restricts the accessible values of $\gam_0$ and $\bar h$. We assume
that the energy $\mean{\Hr}_0$ required to constrain the fluctuations is the
dominant contribution to $\mean{\Delta\Ha}$. The equal sign holds if the
interaction energy $\mean{\Hi}$ is exactly balanced by
$\mean{\Hr}-\mean{\Hr}_0$ describing the energy required to bend (and, in
principle, stretch) the membrane around the adhesion centers. Putting together
Eqs.~\eqref{eq:Z:eff}-\eqref{eq:H:up} we see that the resulting upper bound
$F_N(\cor;\{\x_i\})$ is a function of $\cor$ alone and independent of
interaction details between membrane and substrate, which is our first main
result. In the following we use the minimum $F_N(\cor^\ast)$ as an
approximation to the free energy. We split $F_N=Nf+\Fex_N$ into an ideal part
$f$ and an excess free energy $\Fex_N$. Introducing an effective size $a$ of
adhesion centers with area fraction $\phi\equiv Na^2/L^2$ the ideal part reads
\begin{equation}
  \label{eq:Fid}
  \beta f(\cor) \equiv \frac{(a/\cor)^2}{16\phi} + \ln(\cor/a) + 
  \ln(a/\xi_0)
\end{equation}
with $\xi_0\equiv\sqrt{8\beta\kap}\ell$. The excess free energy becomes
\begin{equation}
  \label{eq:Fex}
  \beta\Fex_N(\cor;\{\x_i\}) \equiv \frac{1}{2}\ln\det\chi
\end{equation}
due to the membrane-induced interactions between adhesion sites. These
interactions are encoded in the matrix $\chi$ with elements
\begin{equation}
  \label{eq:10}
  \chi_{ij}(\cor;\{\x_i\}) =
  \begin{cases}
    1 & i=j, \\
    u(|\x_i-\x_j|/\cor) & i\neq j.
  \end{cases}
\end{equation}
Using $\ell<\rgh\leqslant\rghd$ we find the meaningful range
\begin{equation}
  \label{eq:3}
  \xi_0 < \cor \leqslant \frac{1}{\sqrt{2\pi^3}} L
  \simeq 0.13 L
\end{equation}
for the decay length $\cor$ of the effective interactions. For large $\cor$ we
expect a crossover to free fluctuations while below $\xi_0$ the membrane is
effectively so stiff that a translation of adhesion centers does not require
to bend the membrane. Using Eq.~\eqref{eq:H:up} the temperature and bending
rigidity $\beta\kap$ only set this crossover length and do not influence the
value $\cor^\ast$ that minimizes $F_N$.

\begin{figure}[t]
  \centering
  \includegraphics{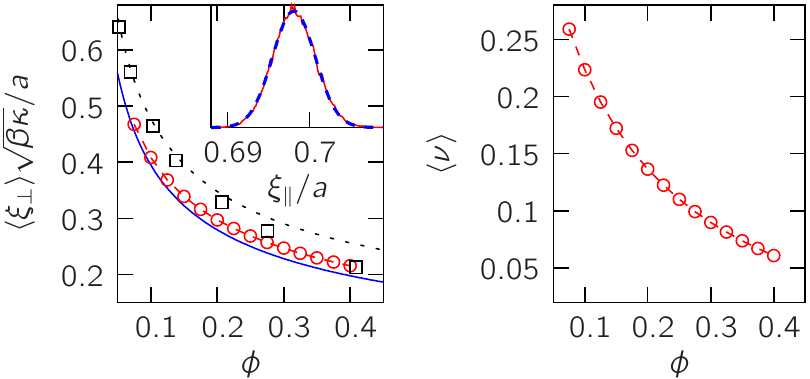}
  \caption{(color online) Left panel: the mean thermal roughness $\mean{\rgh}$
    \textit{vs}. area fraction $\phi$ for the lattice gas (circles) and the
    non-interacting adhesion centers [Eq.~\eqref{eq:non}, solid line]. For
    comparision the \textit{global} thermal roughness reported in
    Ref.~\cite{krob07} is show (squares) with fit
    $\grgh\simeq0.14\phi^{-1/2}+0.035$ (dashed line). Inset: probability
    distribution (solid line) of $\cor$ for the lattice gas ($\phi=0.3$),
    which is well described by a Gaussian (dashed line). Right panel: the mean
    effective interaction strength $\mean{\nu}$. The standard deviation of
    $\nu$ is smaller than symbols.}
  \label{fig:mean}
\end{figure}

We first consider non-interacting adhesion sites, i.e., the off-diagonal
elements of $\chi$ vanish and $\Fex_N=0$. Minimizing $f(\cor)$ leads to the
result
\begin{equation}
  \label{eq:non}
  \cor^\ast = \frac{a}{\sqrt{8\phi}}, \qquad
  \rgh^\ast = \frac{1}{8\sqrt\phi}\frac{a}{\sqrt{\beta\kap}}.
\end{equation}
Such a scaling of the thermal roughness with the inverse square root of the
area fraction has been found in computer simulations~\cite{krob07}. Moreover,
the prefactor $\simeq0.14$ for $\grgh$ determined numerically in
Ref.~\cite{krob07} compares favorably with our $1/8$, see also
Fig.~\ref{fig:mean}. Of course, the divergence for $\phi\ra0$ is only apparent
since we expect a crossover to free fluctuations with $\rgh\simeq\rghd$ if the
membrane is pinned by only a few adhesion centers.


In general $\cor^\ast=\cor^\ast(\{\x_i\})$ will depend on the geometry
of adhesion sites and, therefore, is a fluctuating quantity for mobile
sites, i.e., ligands are able to move in the substrate (e.g., through
using a supported lipid bilayer as substrate). The effective
interactions between adhesion sites are attractive ($u>0$) and one
expects a clustering of sites. This will lead to a larger thermal
roughness $\rgh$ compared to the non-interacting case. To study the
question whether these attractions can actually lead to condensation
we turn to the simplified case of adhesion centers moving on a square
lattice with $\Ns$ sites and lattice constant $a$. The area fraction
is $\phi=N/\Ns$. We define a symmetric $N\times N$ link matrix $b$
with entries $b_{ij}=1$ if the two centers $i$ and $j$ are neighbors
and $b_{ij}=0$ otherwise. Then $\chi_{ij}=\delta_{ij}+u_1 b_{ij}$ with
$u_1\equiv u(a/\cor)<1$. The number of links is
$\Nl=(1/2)\sum_{ij}b_{ij}$. Expanding the excess free energy
Eq.~\eqref{eq:Fex} to second order in $u_1$ leads to
$\beta\Fex_N\approx-\nu\Nl$ with interaction strength $\nu\equiv
u_1^2/2$. This approximation implies only nearest neighbor
interactions and holds for $\cor<a$. Hence, the statistics of the
adhesion centers can be inferred from a two-dimensional lattice gas
with constant density (or, equivalently, the Ising model with constant
magnetization) and Kawasaki diffusion dynamics. This mapping
constitutes the second main result of this Rapid
Communication. However, in contrast to the standard lattice gas the
interaction strength $\nu$ is not constant but depends, as does
$\cor$, on the number of links $\Nl$.

\begin{figure}[t]
  \centering
  \includegraphics{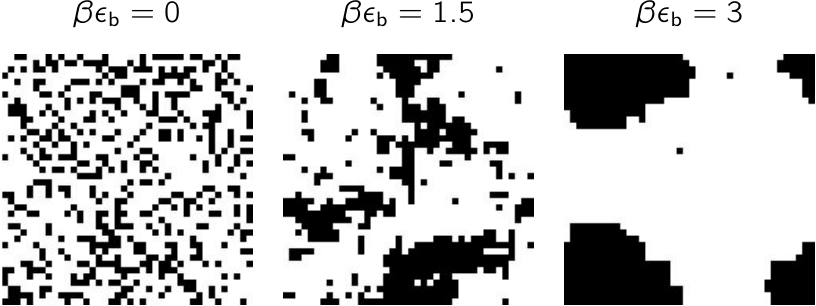}
  \caption{Three snapshots of the lattice gas (adhesion sites are black
    squares) for area fraction $\phi=0.3$ and different values of the binding
    energy $\eb$. The snapshots have been taken after $10^5$ sweeps starting
    from random configurations.}
  \label{fig:snapshot}
\end{figure}

We have performed Monte Carlo simulations of the lattice gas with
$\Ns=40^2$. In Fig.~\ref{fig:mean} we show the averaged thermal roughness
$\mean{\rgh}$ as a function of the area fraction $\phi$ for both the lattice
gas and the ideal non-interacting case [Eq.~\eqref{eq:non}]. For the lattice
gas the decay length $\cor$ is larger due to the clustering of adhesion
centers, which implies that membrane regions can move further away from the
substrate exhibiting a larger thermal roughness as compared to the
non-interacting case. The decay length fluctuates with a narrow distribution
that is Gaussian to a very good degree. For comparison the thermal roughness
$\grgh$ as obtained numerically in Ref.~\cite{krob07} is plotted. Also shown
in Fig.~\ref{fig:mean} is the interaction strength $\nu$ \textit{vs.} the area
fraction. Since for a larger number of adhesion centers the membrane is pulled
closer to the substrate the thermal roughness, and therefore the interaction
strength, decrease. The effective interaction strength is purely entropic as
one would of course expect. Although the interaction strength is not constant
its distribution is rather narrow and hence we compare its mean value
$\mean{\nu}$ to the critical value $\nu_\text{c}\approx1/2.27\approx0.44$ of
the standard Ising model. We see that $\mean{\nu}<\nu_\text{c}$ except for
very small area fractions for which we expect the crossover to unconfined
fluctuations. The fact that the simulations show no condensation (see
Fig.~\ref{fig:snapshot} left panel) is therefore due to the fact that the
interaction strength is too small. This situation changes if we allow for
short-ranged interactions with a binding energy $\eb$ between ligands or
receptors or both. Such a binding energy might again be entropic in nature,
e.g., due to hydrophobic mismatch~\cite{bote06} or different types of
receptors with different lengths~\cite{rozy10}. For large enough
$\nu=u_1^2/2+\beta\eb$ we then observe coarsening and condensation of bonds
into large clusters, see Fig.~\ref{fig:snapshot}.


In summary, we have developed an analytic theory for the statistics of mobile
adhesion centers (single receptor-ligand bonds) in adhered membranes. To this
end we have established an upper bound to the free energy that is independent
of details of the interactions between membrane and substrate. For sites
moving on a discrete lattice the resulting minimized free energy can be mapped
onto a lattice gas with constant density but fluctuating interaction
strength. The mean interaction strength is shown to be well below the critical
value of the Ising model. While we have focused on membrane-substrate
interactions conceptually the same situation arises if the membrane is moving
close to a soft surface such as another cell membrane. Then $h(\x)$ denotes
the inter-membrane distance and $\kap$ is to be replaced by the effective
bending rigidity $\kap=\kap_1\kap_2/(\kap_1+\kap_2)$, where $\kap_i$ are the
bending rigidities of the two membranes.

I thank Oded Farago for helpful discussions and a critical reading of the
manuscript. Financial support by the Alexander-von-Humboldt foundation and by
the Director, Office of Science, Office of Basic Energy Sciences, Materials
Sciences and Engineering Division and Chemical Sciences, Geosciences, and
Biosciences Division of the U.S. Department of Energy under Contract
No.~DE-AC02-05CH11231 is gratefully acknowledged.


\end{document}